**Title**

# Green pathway of Urea Synthesis through Plasma-Ice Interaction: Optimization and Mechanistic Insights with $N_2$ + $CO_2$ and $NH_3$ + $CO_2$ Gas Mixtures

**Authors**

Vikas Rathore*[1], Vyom Desai[1,2], Nirav I. Jamnapara[1,2], Sudhir Kumar Nema[1,2]

**Affiliations**

[1]Atmospheric Plasma Division, Institute for Plasma Research (IPR), Gandhinagar, Gujarat382428, India

[2]Homi Bhabha National Institute, Training School Complex, Anushaktinagar, Mumbai,400094, India

*Author to whom correspondence should be addressed:*

Vikas Rathore

***Email:** vikas.rathore@ipr.res.in, vikasrathore9076@gmail.com

**Abstract**

This study explores a green pathway for urea synthesis using plasma-ice interaction with gas mixtures of $N_2$ + $CO_2$ and $NH_3$ + $CO_2$. Electrical and optical emission spectroscopy were employed to characterize the plasmas, revealing that urea formation involves complex reactions driven by high-energy species, producing reactive nitrogen and carbon intermediates that further react to form urea.

Physicochemical analyses of plasma-treated ice showed increased pH, electrical conductivity (EC), and reduced oxidation-reduction potential (ORP). Optimization of plasma process parameters (gas pressure, applied voltage, and treatment time) was performed to enhance urea formation. Among these parameters, plasma treatment time had the most substantial influence. Increasing treatment time from 20 to 60 minutes significantly impacted physicochemical properties: for $N_2$ + $CO_2$ plasma, pH increased by 21.05%, EC by 184.7%, and ORP decreased by 27.48%; for $NH_3$ + $CO_2$ plasma, pH increased by 27.37%, EC by 239.05%, and ORP decreased by 72.67%, respectively.

The study shows that $NH_3$ + $CO_2$ plasma produces a significantly higher concentration of urea (7.7 mg $L^{-1}$) compared to $N_2$ + $CO_2$ plasma (0.55 mg $L^{-1}$). This is attributed to the direct availability and reactivity of ammonia, which simplifies reaction pathways and enhances intermediate formation. These findings highlight the potential of plasma-ice interaction as an energy-efficient and environmentally friendly method for urea synthesis, offering a sustainable alternative to conventional processes.

## 1. Introduction

The global demand for fertilizers has been escalating due to the need to enhance agricultural productivity to sustain the ever-growing global population (1, 2). Estimates suggest that the world population will reach 9.7 billion by 2050, necessitating a 70% increase in food production compared to current levels (1). The Food and Agriculture Organization predicted in 2016 that the global demand for nitrogen fertilizers would reach 111,591 thousand metric tonnes by 2022 (2).

Among the various nitrogen-based fertilizers, urea ($NH_2CONH_2$) is predominantly used due to its high nitrogen content, which is essential for plant growth (3, 4). The conventional production of urea typically involves the Haber-Bosch process to synthesize ammonia ($NH_3$) from nitrogen ($N_2$) and hydrogen ($H_2$) (5), followed by the Bosch-Meiser process to convert ammonia and carbon dioxide ($CO_2$) into urea (6). While effective, these processes are extremely energy-intensive, operating at high temperatures and pressures, and heavily reliant on fossil fuels.

The Haber-Bosch process releases a significant amount of greenhouse gases, producing over 2.16 tons of $CO_2$ for every ton of ammonia made (> 30 GJ/ton of ammonia), due to the requirement of high temperature and pressure (7). Approximately 1-2% of the world's energy is consumed in ammonia production (8). The Bosch-Meiser process operates under high pressure (around 150 bar) and elevated temperature (above 180°C), emitting substantial carbon dioxide ($CO_2$) and consuming large amounts of fossil fuels (6). Fertilizer manufacturing is an energy-intensive process, utilizing about 1.2% of the world's total energy consumption, with 93% of this energy going towards nitrogen-based fertilizer manufacturing. Consequently, these processes contribute significantly to greenhouse gas emissions and environmental degradation (6, 9, 10).

Current methods of urea production face several challenges. High energy consumption is a major concern as the Haber-Bosch process requires substantial energy input, making it one of the largest industrial consumers of natural gas (9, 10). The carbon footprint of these production processes is also significant, emitting large quantities of $CO_2$ and exacerbating climate change (6, 7). Furthermore, these methods are resource-intensive, demanding significant capital investment in infrastructure and raw materials (6, 9, 10).

In contrast, plasma technology offers a promising alternative for fertilizer synthesis due to its ability to generate highly reactive species under relatively mild conditions (11-14). Plasma, often described as the fourth state of matter, consists of a mixture of free electrons, ions, radicals, and neutral atoms or molecules. These high-energy species can induce chemical reactions that are not feasible under normal conditions. Plasma-liquid interactions have shown considerable potential in various fields, including microbial inactivation, food preservation, pesticide reduction, liquid fertilizer, seed germination, plant growth, selective killing of cancer cells, wound healing, and wastewater treatment, etc. (15-33) The interaction between plasma and liquids can generate reactive species that facilitate complex chemical reactions, potentially leading to more efficient and environmentally friendly processes (33-36).

Despite the broad potential of plasma technology, its application in urea synthesis remains underexplored. Previous research has primarily focused on plasma applications for nitrogen fixation and ammonia synthesis, but direct urea synthesis using plasma has not been extensively studied. This study aims to address this gap by investigating the feasibility of synthesizing urea through plasma-ice interaction using gas mixtures of $NH_3$ + $CO_2$ and $N_2$ + $CO_2$.

In this study, we propose a novel method for urea synthesis leveraging the high reactivity of plasma species generated from $NH_3$ + $CO_2$ and $N_2$ + $CO_2$ gas mixtures. By exposing ice to these plasmas, we aim to facilitate the formation of reactive nitrogen and carbon intermediates, such as NH, $NH_2$, $NH_2COOH$, and $NH_2COONH_4$, which can further react to form urea. Plasma-assisted catalysis and the synergistic effects of these reactive species are expected to lower the activation energy required for urea formation, making the process more efficient and sustainable.

The manuscript is structured as follows: First, the electrical diagnosis of the plasma is presented to understand the characteristics of the discharge. Next, the various plasma species and radicals are identified using optical emission spectroscopy. The subsequent section explores how the exposure of ice to $NH_3$ + $CO_2$ and $N_2$ + CO2 plasmas alters the physicochemical properties of the ice, including pH, oxidation-reduction potential (ORP), and electrical conductivity (EC). Finally, the methods used to identify and quantify the concentration of urea produced through this plasma-ice interaction are detailed.

## 2. Material and Methods

### 2.1 Experimental Setup

The schematic of urea synthesis by plasma-ice interaction is illustrated in Figure 1. Mixtures of ammonia ($NH_3$) + carbon dioxide ($CO_2$) and nitrogen ($N_2$) + carbon dioxide ($CO_2$) were used as plasma-forming gases for urea production. Ice cubes were prepared by freezing ultrapure Milli-Q water (total dissolved solids: 0 ppm) to eliminate any interference from dissolved species.

A 200 ml volume of ice was placed in an aluminum container inside a vacuum chamber. Both the chamber and container were grounded, while the live electrode was negatively biased. A rotary pump (HHV pump) was used to achieve a base pressure of 0.1 mbar in the chamber. Gas flow rates were controlled using a flow controller and gas dosing valve (needle valve) (Balzer), as depicted in Figure 1. Gas pressure was monitored using a KMV vacuum technology pressure gauge and display.

Plasma was generated using a high voltage pulse DC power supply (2.4 kVA, 10 kHz) with a 60% duty cycle. The discharge gas plasma was characterized electrically by monitoring the electrode voltage with a Tektronix P5100 high voltage probe and TDS 2014C oscilloscope. Discharge current was measured using an Ion Physics Corporation CM-100-MG Rogowski coil and oscilloscope.

Optical emission spectroscopy was employed to analyze the emission spectra of the plasma from 200 to 600 nm using a StellarNet Inc EPP2000-UV spectrometer and optical fiber.

**2.2 Physicochemical Properties and Urea Concentration Measurement**

The physicochemical properties of plasma-treated ice were assessed by melting the ice and bringing the liquid to ambient temperature (~24 °C). Measurements included pH, oxidation-reduction potential (ORP), electrical conductivity (EC), and refractive index (RI), determined using specific instruments: a pH meter (Eutech pH 700 with pH electrode), an ORP meter (Eutech pH 700 with ORP electrode), an EC meter (Contech Instrument Ltd.), and a digital refractometer (HM Digital Pvt. Ltd.), respectively.

Urea concentration in the liquid was quantified colorimetrically using the diacetyl monoxime method. Freshly prepared reagents were utilized, and unknown urea concentrations were deduced from a standard curve generated with urea solutions ranging from 0 to 10 mg $L^{-1}$.

Two reagents, acid reagent and color reagent, were freshly prepared to measure urea concentration. The color reagent was made by dissolving 134 mg of diacetyl monoxime and 33.5 mg of thiosemicarbazide in 100 ml of ultrapure Milli-Q water. The acid reagent was prepared by dissolving 25 mg of ferric chloride hexahydrate in 0.5 ml of 85% phosphoric acid and 750 µl of ultrapure Milli-Q water, followed by adding 62.5 µl of this solution to 25 ml of 98% sulfuric acid and 100 ml of ultrapure Milli-Q water, then stirring well to obtain the final acid reagent.

Equal volumes of the color reagent, acid reagent, and known/unknown urea samples (300 µl each) were mixed and heated in a boiling water bath for 30 minutes. The samples were then cooled to room temperature, and absorbance was recorded at 525 nm using a UV-Vis spectrophotometer (SHIMADZU UV-2600). All chemicals used were of high purity and laboratory grade.

### 2.3 Data Analysis

Experimental results were obtained with at least three replications and are presented as mean ± standard deviation ($\mu \pm \sigma$). Statistical analysis utilized analysis of variance (ANOVA), followed by Fisher's LSD (Least Significant Difference) post-hoc test to assess significant differences between groups.

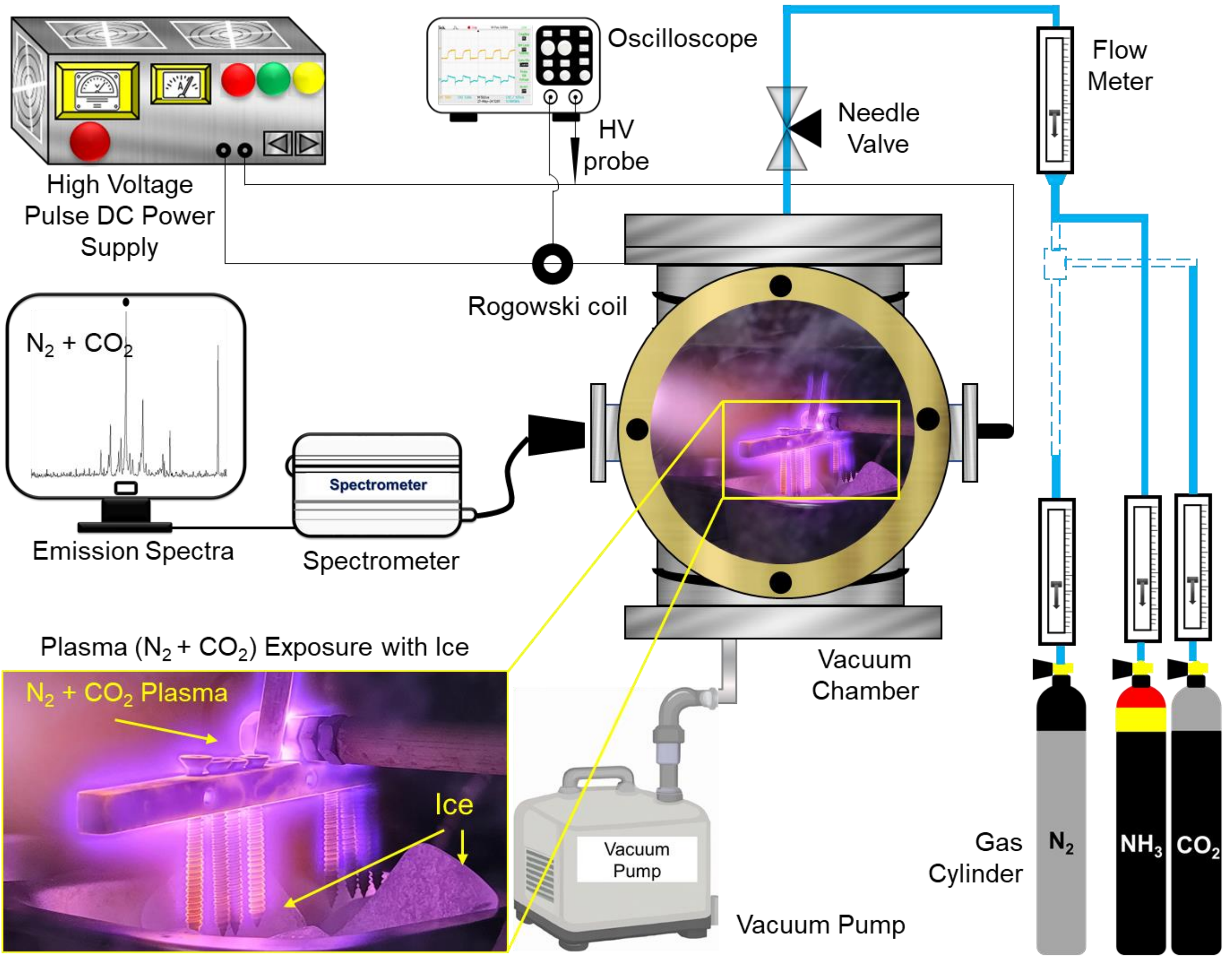

Figure 1. Schematic of urea production via plasma-ice interaction using $N_2$ + $CO_2$ or $NH_3$ + $CO_2$ gas mixtures, including plasma diagnosis through electrical and optical emission diagnostics.

## 3. Results and Discussion

### 3.1 Electrical Characterization of Plasma

The voltage and current waveforms for $NH_3$ + $CO_2$ plasma and $N_2$ + $CO_2$ plasma are depicted in Figure 2. These waveforms illustrate the behavior of the plasma, with the current waveform reflecting the flow of electrons, radicals, and species through the discharge gases. When a negative periodic pulse DC (direct current) is applied across the $NH_3$ + $CO_2$ or $N_2$ + $CO_2$ gas, gas discharge occurs, evidenced by a sudden rise in the negative discharge current (37) as shown in Figure 2.

The observed peak-to-peak current for $NH_3$ + $CO_2$ plasma and $N_2$ + $CO_2$ plasma at the same applied voltage were 2.12 A and 3 A, respectively. This higher discharge current in $N_2$ + $CO_2$ plasma is further supported by emission spectroscopy results, which show that the intensity of observed peaks in $N_2$ + $CO_2$ plasma is about eight times higher than in $NH_3$ + $CO_2$ plasma (Figure 3). The proposed reason for this is the higher electron density in $N_2$ + $CO_2$ plasma, leading to a higher peak-to-peak current. $N_2$, being a more stable diatomic molecule, allows for more straightforward breakdown and subsequent reactions compared to $NH_3$, which is a polar molecule with lone pairs of electrons on the nitrogen atom. This results in more efficient ionization and higher current in $N_2$ + $CO_2$ plasma.

Additionally, variations were observed in both the voltage and current waveforms of $NH_3$ + $CO_2$ and $N_2$ + $CO_2$ plasmas. $NH_3$ + $CO_2$ plasma exhibited greater fluctuations in the voltage waveform and higher voltage peaks at the same applied voltage compared to $N_2$ + $CO_2$ plasma. Furthermore, $N_2$ + $CO_2$ plasma showed a rise and fall in current during the current rising phase, which was absent in $NH_3$ + $CO_2$ plasma. These variations are attributed to the complex plasma chemistry of the generated radicals and species during the interactions of $NH_3$ + $CO_2$ and $N_2$ + $CO_2$.

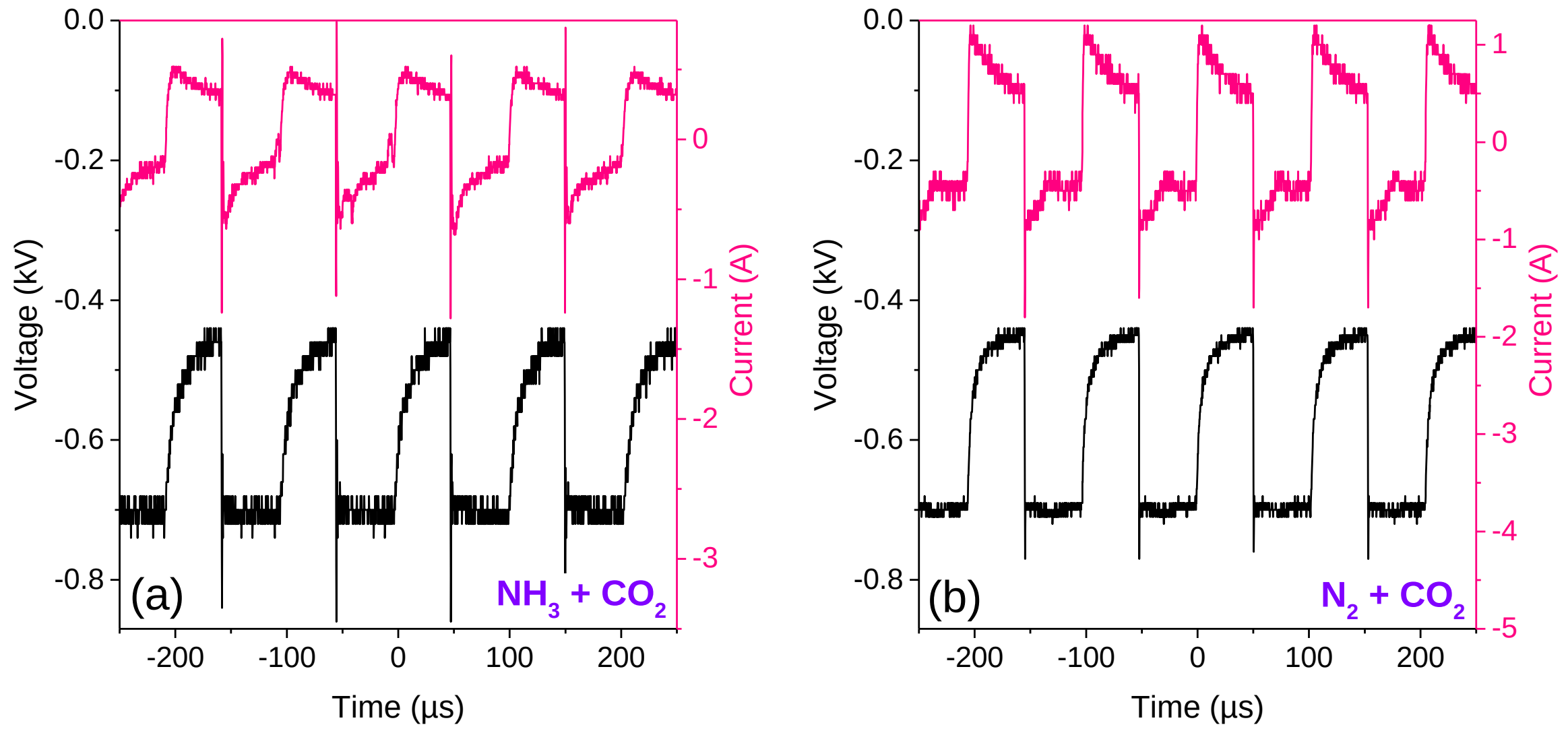

Figure 2. Voltage-current waveforms of (a) $NH_3$ + $CO_2$ plasma and (b) $N_2$ + $CO_2$ plasma during plasma-ice interaction.

### 3.2 Optical Emission Characterization of Plasma

The emission spectra of $NH_3$ + $CO_2$ or $N_2$ + $CO_2$ plasmas observed during plasma-ice interaction, ranging from 200 to 600 nm, are shown in Figure 3. The observed intensity of $N_2$ + $CO_2$ plasma was substantially higher compared to $NH_3$ + $CO_2$ plasma under the same operating parameters, such as gas pressure and applied voltage. Both emission spectra primarily display emission band peaks of the $N_2$ second positive system (SPS) ($C\ ^3\Pi_u \rightarrow B\ ^3\Pi_g$), $N_2^+$ first negative system (FNS) ($B\ ^2\Sigma_u^+ \rightarrow X\ ^2\Sigma_g^+$) and $CO^+$ ($A\ ^2\Pi \rightarrow X\ ^2\Sigma$), etc. transition (38-44). Among these, the emission band peaks of $N_2^+$ FNS are significantly more intense.

In both $N_2$ + $CO_2$ (Figure 3(a)) and $NH_3$ + $CO_2$ (Figure 3(b)) plasmas, the formation of these excited states and transitions occurs through processes such as electron impact excitation, ionization, and recombination. These transitions result in the emission of characteristic spectral lines as shown in Figure 3.

In $N_2 + CO_2$ or $NH_3 + CO_2$ plasma, the formation of $N_2$ SPS occurs when accelerated electrons, influence by of the applied high voltage, collide with $N_2$ molecules, ionizing them to produce $N_2^+$ ions and free electrons. These electrons then excite $N_2$ molecules from the ground state to an excited state. The subsequent decay of these excited molecules (C $^3\Pi_u$) back to the B $^3\Pi_g$ state results in the emission of characteristic photons of the $N_2$ SPS as shown in Figure 3.

$N_2^+$ ions and $CO^+$ are formed in the plasma through processes such as electron impact ionization of $N_2$ and $CO_2$ molecules. These ions can then undergo transitions from the excited state ($N_2^+$ (B $^2\Sigma_u^+$) or $CO^+$ (A $^2\Pi$)) to the ground state ($N_2^+$ (X $^2\Sigma_g^+$) or $CO^+$ (X $^2\Sigma$)), emitting visible photons characteristic of the $N_2^+$ FNS and $CO^+$.

The role of these species and radicals in the formation of urea after plasma-ice interaction will be discussed in the upcoming results and discussion section.

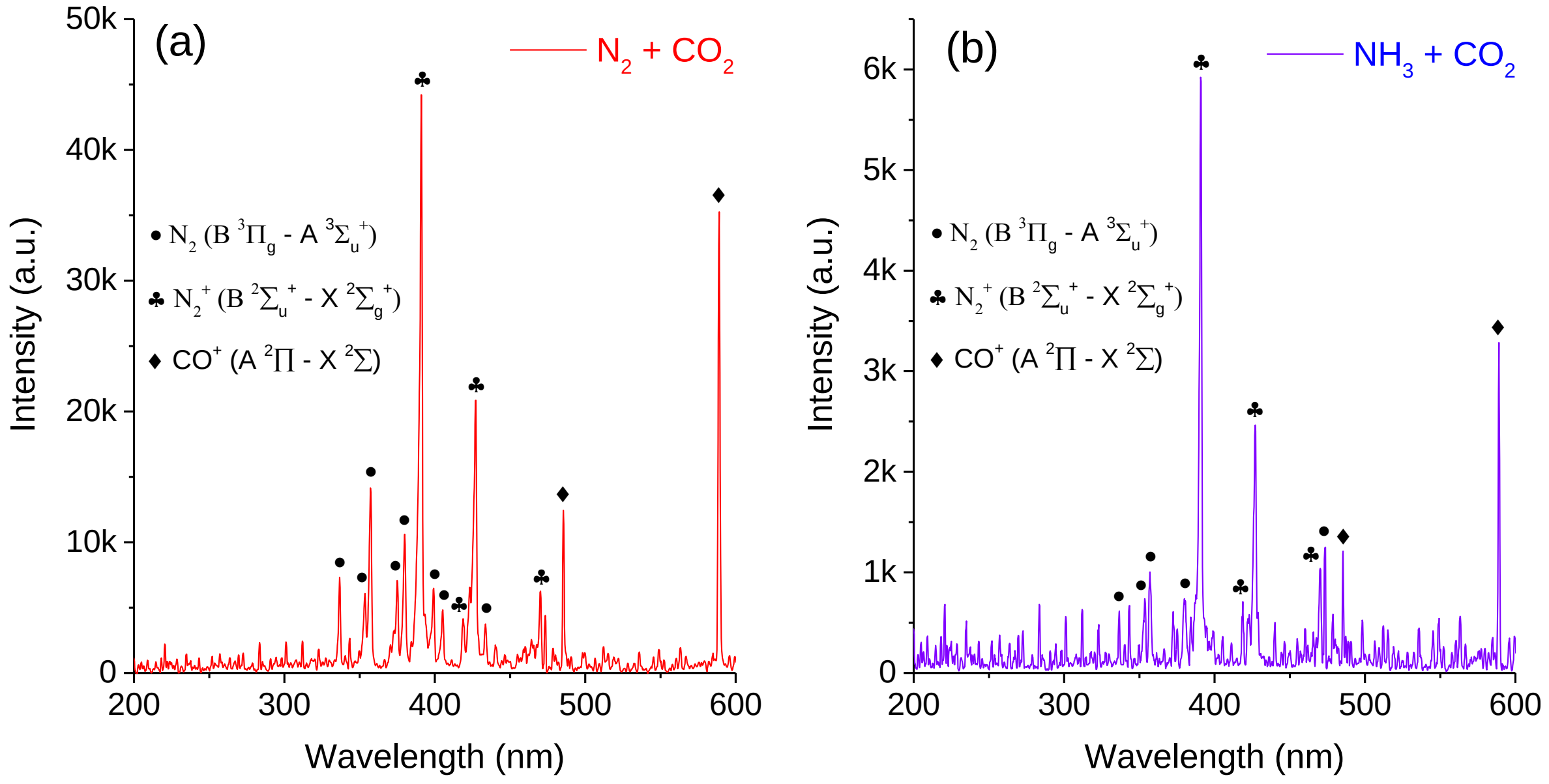

Figure 3. Emission spectra of (a) $NH_3$ + $CO_2$ plasma and (b) $N_2$ + $CO_2$ plasma during plasma-ice interaction.

### 3.3 Physicochemical Properties of Plasma-Treated Ice

The physicochemical properties of ice treated with $N_2$ + $CO_2$ or $NH_3$ + $CO_2$ plasma are detailed in Figure 4. Measurements were performed on the melted ice at room temperature, focusing on pH, oxidation-reduction potential (ORP), and electrical conductivity (EC). Plasma treatment generates numerous reactive species within the ice, necessitating the measurement of these properties to identify the physicochemical traits of the dissolved species.

Figure 4 illustrates that plasma treatment with $N_2$ + $CO_2$ or $NH_3$ + $CO_2$ results in an increase in both pH and EC, and a decrease in ORP. The reduction in ORP indicates a higher concentration of reducing species formed in the ice post-treatment. The rise in pH suggests the formation and absorption of basic species within the ice such as hydroxide ion as shown in equations (5-7). The enhanced EC signifies the presence of ions formed and absorbed during plasma treatment, contributing to increased conductivity.

The influence of various process parameters, such as gas pressure, applied voltage, and plasma treatment time, on these physicochemical properties is also presented in Figure 4. All process parameters positively impact the enhancement of the physicochemical properties of ice, leading to increased pH and EC and decreased ORP. Among these parameters, plasma treatment time exhibits a significantly higher impact compared to others.

Specifically, increasing the gas pressure from 0.3 mbar to 0.5 mbar, while keeping the applied voltage constant at 500 V and treatment time at 20 minutes, results in the following changes (Figure 4 (a, d, g)): for $N_2$ + $CO_2$ plasma, pH increases by 1.83%, EC by 38.23%, and ORP decreases by 4.81%; for $NH_3$ + $CO_2$ plasma, pH increases by 5.33%, EC by 23.64%, and ORP decreases by 6.67%.

After optimizing the gas pressure to 0.5 mbar, the effect of increasing applied voltage was investigated, maintaining the gas pressure at 0.5 mbar and treatment time at 20 minutes. Increasing the applied voltage from 500 V to 600 V results in the following changes (Figure 4 (b, e, h)): for $N_2$ + $CO_2$ plasma, pH increases by 5.12%, EC by 53.2%, and ORP decreases by 13.62%; for $NH_3$ + $CO_2$ plasma, pH increases by 20.25%, EC by 54.41%, and ORP decreases by 28.57%.

Similarly, with the optimized applied voltage at 600 V, the effect of increasing plasma treatment time was investigated, keeping the gas pressure at 0.5 mbar and applied voltage at 600 V. Increasing the plasma treatment time from 20 minutes to 60 minutes results in the following changes (Figure 4 (c, f, i)): for N2 + CO2 plasma, pH increases by 21.05%, EC by 184.7%, and ORP decreases by 27.48%; for NH3 + CO2 plasma, pH increases by 27.37%, EC by 239.05%, and ORP decreases by 72.67%.

These observations highlight the substantial effects of gas pressure, applied voltage, and plasma treatment time on the physicochemical properties of plasma-treated ice. Adjustments in these process parameters lead to significant modifications in pH, EC, and ORP, emphasizing their critical roles in optimizing plasma treatment outcomes.

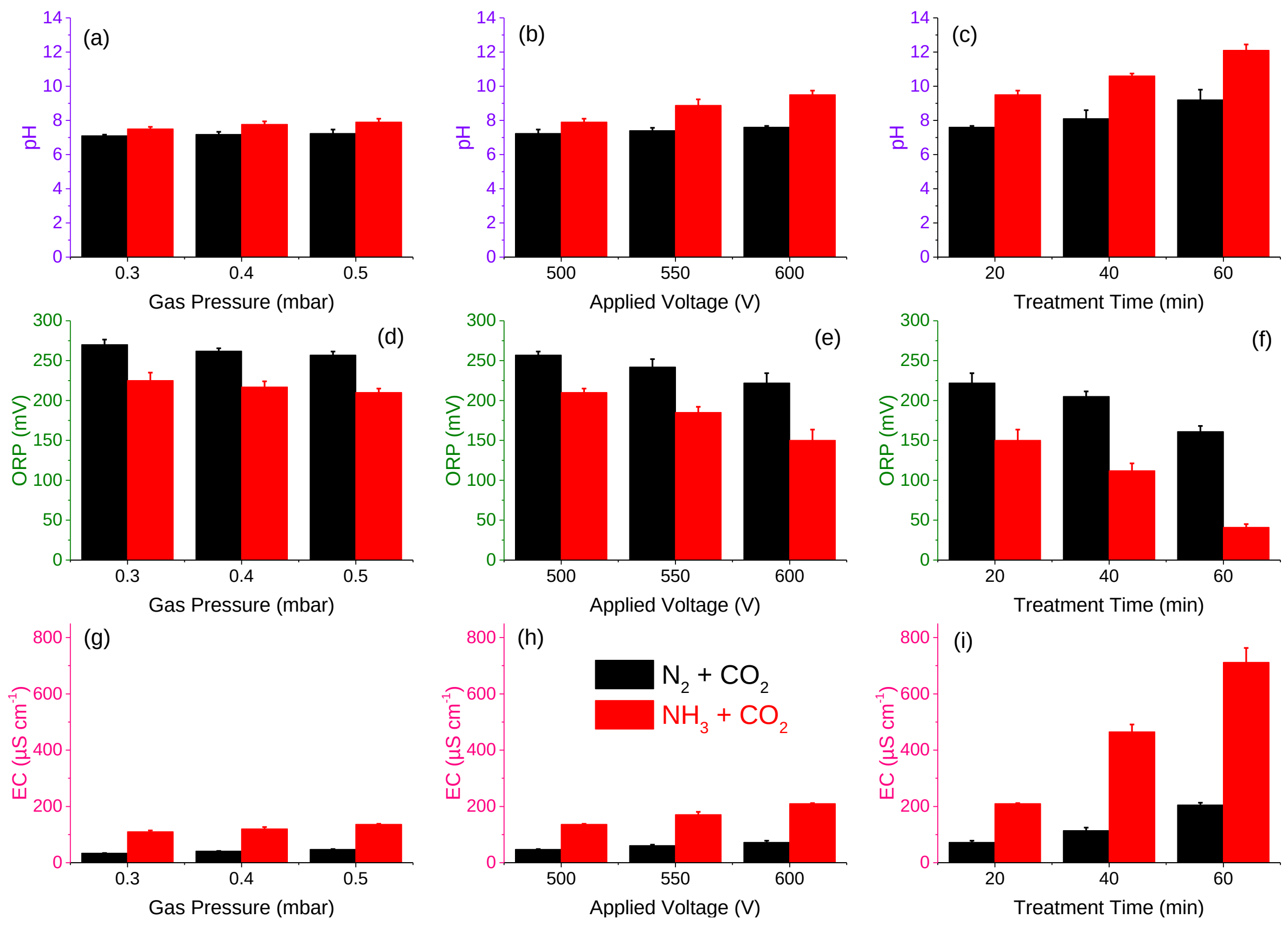

Figure 4. Physicochemical properties plasma treated ice (liquid phase): (a-c) pH, (d-f) Oxidation-Reduction Potential (ORP), and (g-i) Electrical Conductivity (EC), under varying gas pressure, applied voltage, and plasma treatment time.

## 3.4 Urea Formation and Concentration Measurement

As discussed earlier, plasma treatment of ice generates numerous reactive species within the ice. This investigation evaluates the feasibility of urea formation after plasma exposure using $N_2 + CO_2$ or $NH_3 + CO_2$ plasmas. The precursors for urea formation — nitrogen, hydrogen, and carbon — are provided by $N_2$, $CO_2$, and $H_2O$ when using the $N_2$ + $CO_2$ gas mixture, and by $NH_3$, $CO_2$, and $H_2O$ when using the $NH_3$ + $CO_2$ gas mixture.

The measured concentration of urea formed after plasma exposure with $N_2$ + $CO_2$ or $NH_3$ + $CO_2$ as the plasma-forming gas is shown in Figure 5 (a-b). The formation and dissolution of urea in ice also change its refractive index, as measured by a refractometer (Figure 5 (c-e)). Similar to physicochemical properties, increasing plasma treatment time significantly enhances the urea concentration compared to other process parameters. The maximum observed concentrations of urea were 7.7 mg $L^{-1}$ for $NH_3$ + $CO_2$ plasma and 0.55 mg $L^{-1}$ for $N_2$ + $CO_2$ plasma. This higher concentration is also reflected in the refractive index (RI) measurements, with a 200% higher RI observed when $NH_3$ + $CO_2$ was used compared to $N_2$ + $CO_2$ (Figure 5 (f)). The optimal process parameters for these maximum values were 0.5 mbar gas pressure, 600 V applied voltage, and 60 minutes of plasma treatment time.

The observed maximum concentration of urea produced using $NH_3$ + $CO_2$ plasma was 1300% higher than that produced with $N_2$ + $CO_2$ plasma. This significant difference can be attributed to the underlying reaction mechanisms, the availability of reactants, and the efficiency of intermediate formation in the plasma environment.

In the $NH_3$ + $CO_2$ plasma, ammonia ($NH_3$) is already present as a primary reactant, directly providing $NH_2$ groups necessary for urea formation without additional nitrogen fixation steps. The proposed reaction mechanism suggests that $NH_3$ can readily react with $CO_2$ to form intermediates such as carbamic acid (NH2COOH) as shown in equation (9) (45), a crucial precursor to urea.

The presence of $NH_3$ facilitates the direct formation of carbamic acid ($NH_2COOH$) when it reacts with $CO_2$. This intermediate is essential for urea synthesis. The $NH_3$ + $CO_2$ plasma enables more straightforward reaction pathways to urea compared to $N_2$ + $CO_2$ plasma, where nitrogen must first be fixed and then react with hydrogen or other intermediates to form ammonia or other nitrogenous compounds (46). The direct use of ammonia in the reaction bypasses the energy-intensive nitrogen fixation steps, leading to a higher overall yield of urea.

When subjected to plasma, $NH_3$ molecules can be easily ionized and excited, forming NH and $NH_2$ radicals and other reactive species (H, $H_2$, etc.) (equation (1)) that efficiently participate in the urea synthesis. The emission band peaks of NH (A $^3\Pi_i \rightarrow$ X $^3\Sigma^-$) (0 - 0) overlap with N2 SPS (C $^3\Pi_u \rightarrow$ B $^3\Pi_g$) (0 - 0), making them difficult to identify in the emission spectra of $NH_3$ + $CO_2$ plasma (Figure 3 (b)) (47).

In $N_2$ + $CO_2$ plasma, nitrogen molecules ($N_2$) need to be dissociated and fixed into the ice as reactive nitrogen species like NH or $NH_2$ (48, 49) (equations (2-6)) before they can react with $CO_2$ to form intermediates such as carbamic acid or ammonium carbamate (equations (9, 10)). This additional nitrogen fixation step in $N_2$ + $CO_2$ plasma is less efficient, more energy-consuming, and may require a catalyst to accelerate the reactions, resulting in a lower overall yield of urea compared to the direct use of $NH_3$ in $NH_3$ + $CO_2$ plasma.

$NH_3$ + $CO_2$ plasma likely forms more stable and readily convertible intermediates leading to urea, such as ammonium carbamate ($NH_2COONH_4$) (equation (10)) (45). The intermediates formed in $NH_3$ + $CO_2$ plasma are more efficiently converted to urea due to the more direct and fewer reaction steps involved. The dehydration of ammonium carbamate ($NH_2COONH_4$) yields urea ($NH_2CONH_2$) (equation (11)) (45).

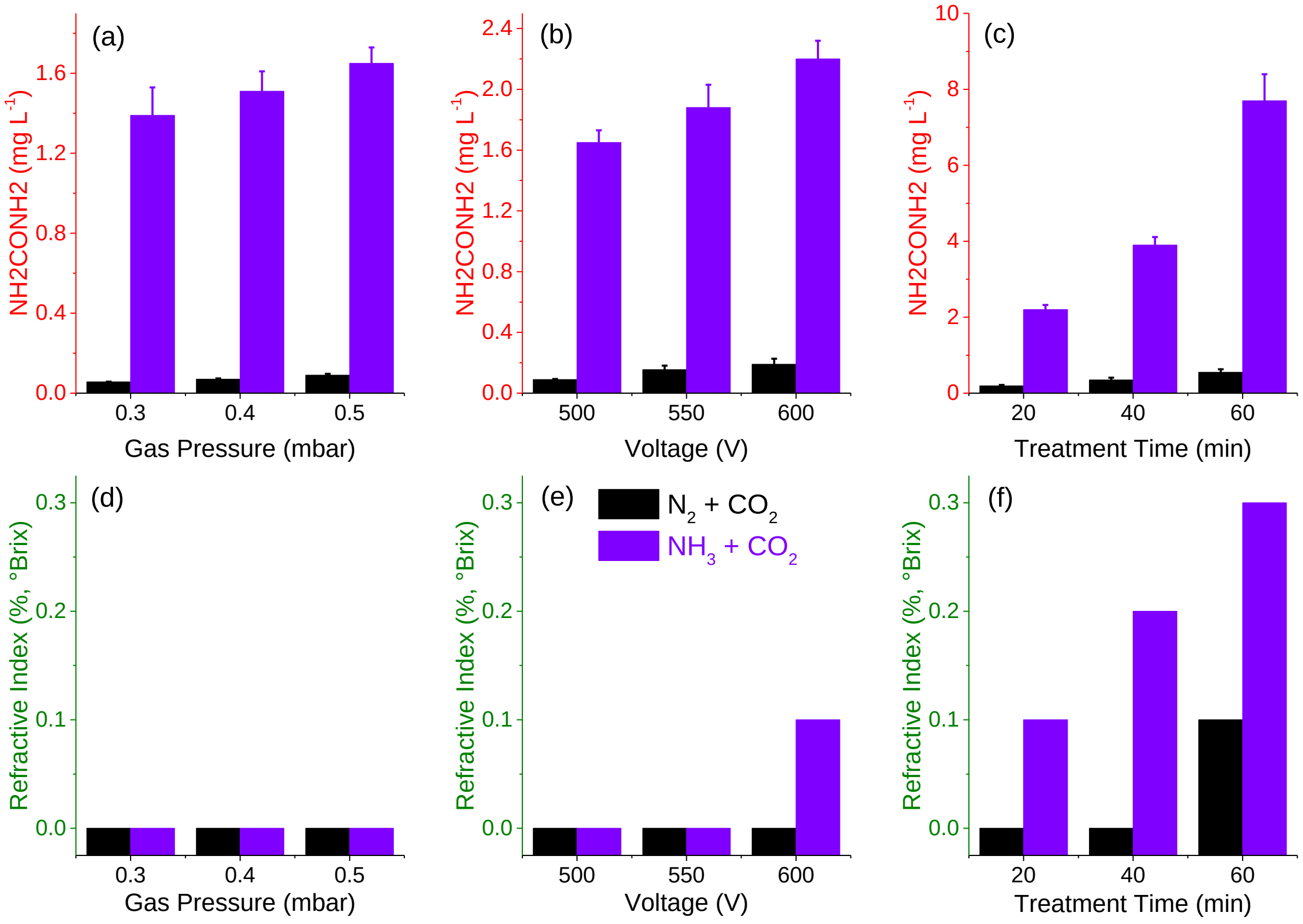

Figure 5. (a-c) Urea ($NH_2CONH_2$) concentration and (d-f) refractive index of liquid (melted ice) after plasma-ice exposure under varying gas pressure, applied voltage, and plasma treatment time.

## 4. Discussion

The formation of urea ($NH_2CONH_2$) using $N_2$ + $CO_2$ or $NH_3$ + $CO_2$ plasma exposed to ice involves complex chemical processes driven by the high-energy species generated in the plasma, such as $N_2$ Second Positive System (SPS), $N_2^+$ First Negative System (FNS), and $CO^+$ (Figure 3) (22, 36, 38). These excited states and transitions form through electron impact excitation, ionization, and recombination processes. In $N_2$ + $CO_2$ plasma, accelerated electrons collide with $N_2$ molecules, ionizing them to produce $N_2^+$ ions and free electrons, which excite $N_2$ molecules (40-42, 44, 50). The subsequent decay of these molecules emits SPS photons. Similarly, $N_2^+$ and $CO^+$ ions form through electron impact ionization of $N_2$ and $CO_2$ molecules, and their transitions emit visible photons characteristic of the $N_2^+$ FNS and $CO^+$ transitions (51, 52). These high-energy species are crucial in breaking molecular bonds and initiating reactions that lead to urea formation.

The formation of urea ($NH_2CONH_2$) in $N_2$ + $CO_2$ or $NH_3$ + $CO_2$ plasmas exposed to ice involves complex chemical processes driven by high-energy species. In both plasmas, these species react with water to form reactive nitrogen species such as NH, $NH_2$, NOx (NO, $NO_2$, etc.) (48-50, 53). For instance, $N_2$* reacting with $H_2O$ produces NH and OH radicals, while $N_2^+$ interacting with $H_2O$ forms $NH_2$ and OH ions (48, 49). CO2 in the plasma can be ionized to form $CO^+$ and $CO_2^+$ ions (38, 51, 52), which further react with water to produce reactive carbon species like $COOH^+$ (45, 46). The proposed reaction mechanism for urea formation during plasma-ice interaction is as follows:

$$NH_3 \rightarrow NH_2 + H \; or \; (NH,\; H_2,\; H,\; etc.) \quad (1)$$

$$H_2O \rightarrow H + OH \rightarrow 2H + O \quad (2)$$

$$N_2 \rightarrow N_2^* \rightarrow 2N \quad (3)$$

$$N + xH \rightarrow NH_x \; (NH, NH_2, NH_3, etc.) \quad (4)$$

$$N^* + H_2O \rightarrow NH + OH \quad (5)$$

$$N^* + 2H_2O \rightarrow NH_2 + 2OH \quad (6)$$

$$CO_2^+ + H_2O \rightarrow COOH^+ + OH \quad (7)$$

$$COOH^+ + NH_2 + e^- \rightarrow NH_2COOH \quad (8)$$

$$NH_3 + CO_2 \rightarrow NH_2COOH \quad (9)$$

$$NH_2COOH + NH_3 \rightarrow NH_2COONH_4 \quad (10)$$

$$NH_2COONH_4 \rightarrow NH_2CONH_2 + H_2O \quad (11)$$

These high-energy species facilitate the formation of carbamic acid ($NH_2COOH$) when $NH_2$ radicals react with $CO_2$. Ammonia ($NH_3$) then reacts with carbamic acid to form ammonium carbamate ($NH_2COONH_4$) (45), which dehydrates under plasma conditions to form urea (45). The high-energy electrons and ions generated in the plasma act as catalysts, lowering the activation energy required for these reactions (51). The simultaneous presence of reactive nitrogen and carbon species increases the likelihood of forming intermediates like carbamic acid and ammonium carbamate, which are precursors to urea (45). Additionally, reactions on the surfaces of plasma-treated ice, where reactive species can adsorb and react more efficiently than in the gas phase, contribute to urea formation.

The significantly higher urea concentration in $NH_3$ + $CO_2$ plasma compared to $N_2$ + $CO_2$ plasma can be explained by the underlying reaction mechanisms and the efficiency of intermediate formation. In $NH_3$ + $CO_2$ plasma, ammonia ($NH_3$) is present as a primary reactant, directly providing $NH_2$ groups necessary for urea formation without additional nitrogen fixation steps. Ammonia reacts readily with $CO_2$ to form carbamic acid, a crucial precursor to urea. The direct formation of carbamic acid and subsequent steps in $NH_3$ + $CO_2$ plasma are more straightforward and efficient than in $N_2$ + $CO_2$ plasma, where nitrogen must first be fixed and then react with hydrogen or other intermediates. This bypasses energy-intensive nitrogen fixation steps, leading to a higher overall yield of urea. The intermediates formed in $NH_3$ + $CO_2$ plasma are more stable and readily convertible to urea, making the synthesis process more efficient compared to $N_2$ + $CO_2$ plasma.

This research not only demonstrates the potential of plasma technology in green chemistry but also lays the groundwork for future studies into sustainable urea production methods. By providing a more efficient and environmentally friendly approach to urea synthesis, this novel method could significantly reduce the environmental impact of fertilizer production, thereby promoting more sustainable agricultural practices.

## 5. Conclusion

The formation of urea in $N_2$ + $CO_2$ or $NH_3$ + $CO_2$ plasma exposed to ice involves a complex interplay of high-energy species generated in the plasma, including $N_2$ SPS, $N_2^+$ FNS, and $CO^+$. These species participate in a series of reactions that produce reactive nitrogen and carbon intermediates, which further react to form urea.

The significantly higher concentration of urea observed with $NH_3$ + $CO_2$ plasma compared to $N_2$ + $CO_2$ plasma is due to the direct availability and higher reactivity of ammonia. This simplifies and enhances the efficiency of the urea synthesis process by facilitating the formation of necessary intermediates more effectively. The $NH_3$ + $CO_2$ plasma avoids the energy-intensive steps of nitrogen fixation required in $N_2$ + $CO_2$ plasma, leading to higher overall yields of urea. This study underscores the potential of using plasma-ice interaction as a green, energy-efficient method for urea synthesis, offering a sustainable alternative to conventional processes.

## Conflict of interests

The authors declare that there are no conflicts of interests.